\documentclass[useAMS,usenatbib]{mn2e}
\usepackage{newtxtext,newtxmath}
\usepackage{multirow}
\usepackage{times}
\usepackage{graphics,epsfig,txfonts}
\usepackage{textcomp}

\usepackage{enumerate}
\usepackage[T1]{fontenc}
\usepackage{ae,aecompl}
\usepackage[usenames, dvipsnames]{color}%added by Vanessa to highlight text in colour
\usepackage[flushleft]{threeparttable}

\usepackage{natbib}

\usepackage{graphicx}	% Including figure files
\usepackage{amsmath}	% Advanced maths commands
\usepackage{amssymb}	% Extra maths symbols
\usepackage{pdflscape}
\title[Radio pulsars in the SMC]{Targeted search for young radio pulsars in the  SMC:  Discovery of two new pulsars}

\author[N. Titus et al.]{
N. Titus$^{1,2}$\thanks{E-mail: naomi@saao.ac.za},
B. W. Stappers$^{3}$,
V. Morello$^{3}$,
M. Caleb$^{3,4,5,6}$,
M. D. Filipovi\'c$^{7}$,
\newauthor
V. A. McBride$^{1,2,8}$,
W. C. G. Ho$^{9,10}$
D. A. H. Buckley$^{2}$
\\
$^{1}$Department of Astronomy, University of Cape Town, Private Bag X3, Rondebosch, 7701, South Africa\\
$^{2}$South African Astronomical Observatory, PO Box 9, Observatory, 7935, South Africa\\
$^{3}$Jodrell Bank Centre for Astrophysics, School of Physics and Astronomy, University of Manchester, Manchester M13 9PL, UK\\
$^{4}$Research School of Astronomy and Astrophysics, Australian National University, ACT, 2611, Australia\\
$^{5}$Centre for Astrophysics and Supercomputing, Swinburne University of Technology, P.O. Box 218, Hawthorn, VIC 3122, Australia\\
$^{6}$ARC Centre of Excellence for All-sky Astrophysics (CAASTRO)\\
$^{7}$Western Sydney University, Locked Bag 1797, Penrith South DC, NSW 1797, Australia\\
$^{8}$IAU Office of Astronomy for Development, Cape Town, South Africa\\
$^{9}$Department of Physics and Astronomy, Haverford College,
370 Lancaster Avenue, Haverford, PA 19041 USA\\
$^{10}$Mathematical Sciences, Physics and Astronomy and STAG Research Centre,
University of Southampton, SO17 1BJ, UK}
\date{Accepted 2019 June 4. Received 2019 June 4; in original form 2019 April 17}

\pubyear{2017}

\begin{document}
\label{firstpage}
\pagerange{\pageref{firstpage}--\pageref{lastpage}}
\maketitle

% Abstract of the paper
\begin{abstract}
We report the first rotation powered pulsars discovered in the Small Magellanic Cloud (SMC) in more than a decade.  \text{PSR~J0043--73} and \text{PSR~J0052--72} were discovered during a Parkes Multi-Beam (PMB) survey of \text{MCSNR~J0127-7332}, and five new, optically selected, supernova remnant (SNR) candidates identified by the XMM Newton survey.  In addition to the candidates, we adjusted the PMB rotation to include an additional nine SNRs and pulsar wind nebulae.  We searched for young pulsars (1\,--\,200\,ms) employing a Fourier analysis with \texttt{PRESTO}, as well as a search for longer period pulsars (200\,ms\,--\,360\,s) with a fast folding algorithm.   Our targeted survey had a limiting flux density of 0.039\,mJy for periods greater than 50\,ms.  Although not the main target of this search it was also sensitive to millisecond pulsars.  \text{PSR~J0043--73} has a period and dispersion measure of 937.42937\,(26)\,ms and 115.1\,(3.4)\,pc\,cm\textsuperscript{-3} respectively, and \text{PSR~J0052--72} has a period of 191.444328\,(46)\,ms and a DM of 158.6\,(1.6)\,pc\,cm\textsuperscript{-3}. 
\end{abstract}

\begin{keywords}
 pulsars: general --  supernova remnants -- Magellanic Clouds
\end{keywords}

%%%%%%%%%%%%%%%%%%%%%%%%%%%%%%%%%%%%%%%%%%%%%%%%%%

%%%%%%%%%%%%%%%%% BODY OF PAPER %%%%%%%%%%%%%%%%%%

\section{Introduction}

The Magellanic Clouds (MCs) are our closest satellite galaxies, and the only galaxies other than the Milky Way in which we have been able to detect radio pulsars.  The MCs have gone through a relatively recent episode of star formation \citep{Antoniou2016}, producing a large number of O and B stars, many of which are companions in high mass X-ray binary (HMXB) systems.  A recent census (\citealt{Haberl2015}) identified 120 HMXBs (of which 63 are X-ray pulsars) in the Small Magellanic Cloud (SMC) -- similar to the population in the Milky Way, despite the fact that the SMC is only one-fiftieth the mass. This is a clear indication that the SMC harbours many neutron stars (NSs) and many in interesting binaries.  However, despite this large population of NSs only five rotation powered pulsars so far have been discovered in the SMC through various MC radio surveys \citep{mcconnell1991,Crawford2001,Manchester2006}.

Massive stars form NSs during core-collapse supernova events.  Most of the newly formed NSs are rotation-powered pulsars \citep{Keane2008}, which emit pulsed, non-thermal radio emission.  Energetic young radio pulsars can produce an outflow of relativistic particles which will interact with their natal SNR and the surrounding interstellar medium, creating a pulsar wind nebula (PWN).  Thus, SNRS and PWNe are excellent candidates for targeted radio pulsar surveys, that being said pulsars have so far only been discovered in $\sim$50\% of SNRs and PWNe \citep{Green2014}.  The lack of SNRs and PWNe associations with pulsars may be the result of beaming effects, or alternatively due to instrumentation sensitivity limits.   \citet{Camilo2003} completed a census of SNR-pulsar associations \citep{KaspiV.M.Helfand2002,Gorham1996,Lorimer1998,Manchester2001} and found that in some cases the pulsars associated with SNRs can be faint.  Hence, deep surveys may be required to detect these young radio pulsars.  

The first of the SMC surveys using the Parkes radio telescope was conducted by \citet{mcconnell1991} with a limiting mean flux density at 610\,MHz of 0.5\,-\,0.8\,mJy for pulsars with periods greater than 500\,ms and dispersion measure (DMs) of 100\,pc\,cm\textsuperscript{-3} or less.  The survey discovered 1 pulsar, \text{PSR~J0045--7319}, later shown to be in a 51 day orbit around a 16\textsuperscript{th} magnitude B star (\citealt{KaspiV.M.JohnstonS.BellJ.F.ManchesterR.N.BailesM.BessellM.LyneA.G.1994}).  In a subsequent SMC survey with the 20\,cm Multi-Beam receiver at Parkes (PMB),  \citet{Crawford2001} discovered \text{PSR~J0113--7220}.  The survey covered a more complete region of the SMC ($\sim$6.7\,deg\textsuperscript{2}) with a limiting flux density of 0.08\,mJy at 1400\,MHz for pulsars with periods greater than 50\,ms and DMs less than 200\,pc\,cm\textsuperscript{-3}.  The most recent and successful SMC survey was conducted by \citet{Manchester2006}, also using the PMB receiver, in which three pulsars were detected (\text{PSR~J0045--7042}, \text{PSR~J0111--7131}, \text{PSR~J0131--7310}).  The advantage of the \citet{Manchester2006} survey was the increased DM range of up to 277\,pc\,cm\textsuperscript{-3}, however it was most sensitive to periods greater than 50\,ms, but with a similar limiting flux density to the \citet{Crawford2001} survey.  

Due to the distance to the SMC ($\sim$60 kpc), the blind surveys only detected the brightest pulsars, since integration time is sacrificed to cover a larger area.  To improve upon the sensitivity of the previous blind surveys we carried out the deepest, most sensitive pulsar searches of five supernova remnant (SNR) candidates, as well as \text{MCSNR~J0127-7332} in the SMC. Finding a young pulsar which is coincident with its natal SNR or within a PWN will enable studies of the initial spin period, velocity, and magnetic field distributions of young NSs, providing insights to the physics governing core-collapse.  It also enable searches at other energies, particularly X-ray observations can facilitate NS cooling studies.  Finding just one new pulsar in the SMC will provide valuable details about the nature of the SMC pulsar population and the progenitor stars that formed it.  Particularly, the low metallicity environment of the SMC will affect mass loss of massive stars uniquely when compared to the Galaxy.  This may be reflected in the distribution of angular momentum of the radio pulsars.

In this paper we present the results for our targeted pulsar searches in the SMC.  In Section 2, we describe the observations, followed by the data reduction process in Section 3.  In Section 4 we present the results, and discuss their implications in Section 5.  Finally, in Section 6, we draw our conclusions.

\section{Source Selection and Observations}

\subsection{Supernova Remnant Candidates}

The XMM-Newton satellite has conducted the deepest complete survey of the SMC in the 0.15--12.0\,keV X-ray band to date \citep{Haberl2012,Sturm2013}. The relatively close and known distance to the SMC (60 kpc) together with the moderate Galactic foreground absorption of NH\,$\approx$\,6$\times$10$^{20}$\,cm\textsuperscript{-2} \citep{Dickey1990} provides an opportunity to study the complete X-ray source population within the SMC. These are dominated by HXMBs, SNRs and super-soft X-ray sources. In total the survey detected 23 SNRs, 20 previously known \citep{Badenes2010} and 3 new candidates, XMMU\,J0049-7306, XMMU\,J0056.6-7208 and XMMU\,J0057.7-7213. 

Following a study of narrow band optical images (H$\alpha$ and SII) and ATCA radio observations (\citealt{Filipovic2005}; \citealt{Payne2007}, \citealt{Roper2015}; Maggi et al. 2019, submitted) of the SMC, nine new SNR candidates were selected for further study. One of these (SNR\,C1) was found to be coincident with one of the newly identified XMM-Newton SNRs \citep{Haberl2012}, namely XMMU J0056.6-7208. Following SALT long slit spectroscopy of 8 of these new candidate SNRs, SII/H$\alpha$ ratios were determined and indicated that 6 were consistent with being SNR. These, plus the two other XMM-Newton SNR candidates were then selected to be observed by SALT using the Fabry-Perot imaging at H$\alpha$, SII and OIII. In the end only 7 SNRs candidates were observed, of which 4 showed strong evidence for SNR shells, namely SNR\,C1 (=XMMU\,J0056.6-7208), SNR\,C3, SNR\,C4 \& SNR\,C9). The 3 others showed either indistinct or small shells (SNR\,C2, SNR\,C6 \& XMMU\,J0057.7-7213). A known SNR, \text{MCSNR\,J0127-7332}, which is associated with a slowly spinning X-ray pulsar SXP\,1062 in a Be X-ray binary \citep{Henault-Brunet2012}, was observed as a control object, which was confirmed as a SNR.

We were awarded 54 hours with the Parkes radio telescope to observe six of the SNR candidates (\text{MCSNR~J0127--7332}, \text{XMMU\,J0049.0--7306}, \text{SNR\,C3}, \text{SNR\,C2}, \text{XMMU\,J0056.6-7208}, \text{SNR\,C9}), which were coincident with X-ray point sources.

\subsection{Parkes Observations}

The PMB receiver \citep{Staveley-Smith1996} has the ability to observe 13 distinct regions on the sky similtaneously, covering a 0.5\,deg\textsuperscript{2} region, making it ideal for surveys.  The PMB consists of the most sensitive central beam, six inner ring beams, and the six least sensitive outer ring beams.  The beam properties are listed in Table~\ref{tab:beams}.

\begin{table}
    \centering
    \caption{The PMB receiver properties as recorded by \citet{Manchester2001}.}
    \begin{tabular}{llll}
    \hline
    \hline
    Beam&Centre&Inner&Outer\\
    \hline
    Gain (K\,Jy\textsuperscript{-1})&0.735&0.690&0.581\\
    Half power beamwidth (arcmin)&14.0&14.1&14.5\\
    Beam ellipticity&0.0&0.03&0.06\\
    Coma lobe (dB)&none&17&14\\
    \hline
    \label{tab:beams}
    \end{tabular}
\end{table}

\begin{figure*}
  \includegraphics[width=\textwidth]{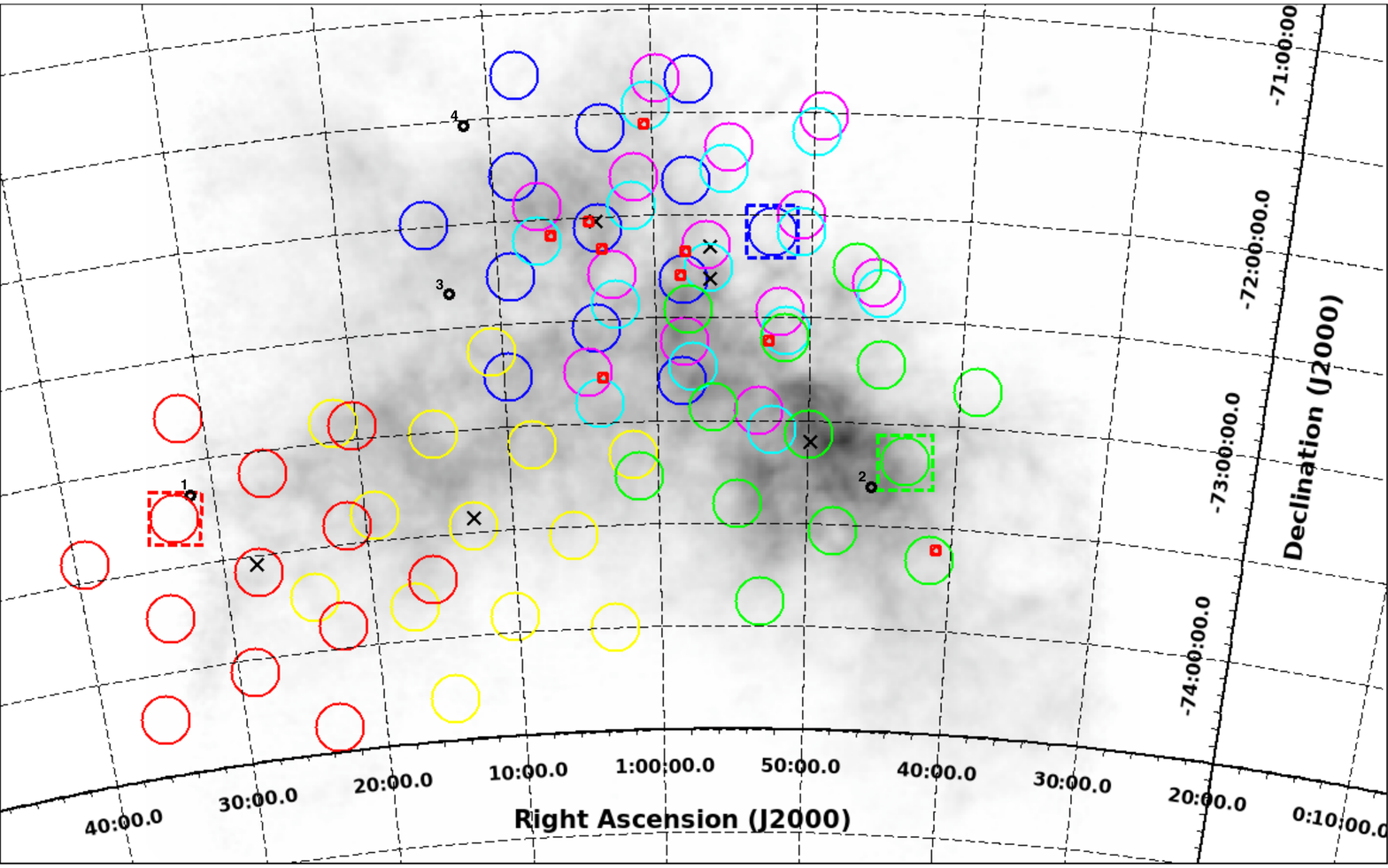}
  \caption{The coloured circles show the PMB orientation, where their 14\,arcmin width is consistent with the half power beam width (HPBW) of the PMB beams  (Table~\ref{tab:beams}), and is superimposed on a HI map of the SMC \citep{Dickey1999a}.  Furthermore, the locations of the five SNR candidates and \text{MCSNR~J0127-7332} are indicated by black crosses, and four of the known SMC radio pulsars are shown by black circles (1: \text{PSR~J0131--7310}; 2: \text{PSR~J0045--7319}; 3: \text{PSR~J0113--7220}; 4:  \text{PSR~J0111--7131}; the fifth known pulsar is off the plot on the right hand corner).  The nine additional SNRs and PWNe are highligted with small red squares.  The dashed squares indicate the beams within which pulsars were detected: \text{PSR~J0131--7310} (red square), the new pulsars, \text{PSR~J0043--73} (green square), and \text{PSR~J0052--72} (blue square).}
  \label{fig:survey}
\end{figure*}

We used the PMB with the Berkeley-Parkes-Swinburne data recorder (BPSR, \citealt{McMahon2008}; \citealt{Keith2010}) to perform deep pulsar searches in the SMC. The focus of the survey was 5 optically selected SNR candidates, as well as a known SNR, \text{MCSNR\,J0127-7332}, which are coincident with X-ray point sources.  The extent of the SNRs was less than the $\sim$14 arcmin half power beam width (HPBW) of the PMB.  The central beam of the PMB was positioned on the SNRs, and then rotated to include an additional 9 known SNRs and PWNe \citep{Filipovic2008a}.  The targets are listed in  Table~\ref{tab:observations}.  One beam of one pointing was adjusted as much as possible to the position of \text{PSR~J0131--7310}, a known radio pulsar in the SMC, with a period and DM of 348.124045581\,(7)\,ms and 205.2\,pc\,cm\textsuperscript{-3} respectively \citep{Manchester2006}.  The PMB beam pattern and configuration is highlighted in Figure~\ref{fig:survey}.

\begin{table*}
\caption{Survey field observed in the SMC with the PMB.}
\label{tab:observations}
\begin{tabular}{llllllllll}
\hline
\hline
Date&Target&Beam&Epoch$^{a}$&RA$_{beam}$$^{b}$&Dec$_{beam}$$^{b}$&$t_{int}$$^c$&Type&Confirmed&Figure~\ref{fig:survey}:\\
&&&(MJD)&(J2000)&(J2000)&(s)&(SNR/PWN)& (yes/no) &PMB pattern\\
\hline
2017--08--25&MCSNR~J0127-7332&1& 57990.50 &01 27 37.66&-73 35 20.50&16\,223& SNR & yes & Red\\
&XMMU\,J0049.0-7306&1& 57990.69 &00 49 54.48&-73 04 09.70&12\,989& SNR & no & Green\\
&DEM\,S5&13& 57990.69 &00 41 24.24&-73 39 24.80&12\,989& PWN & yes & Green\\
&B0050-72.8&3& 57990.69 &00 51 35.18&-72 36 02.30&12\,989& SNR & yes & Green\\
2017--08--28&SNR\,C3&1& 57993.47 &01 03 27.99&-72 03 41.30&16\,217& SNR & no & Blue\\
&IKT\,16&2$^{d}$& 57993.47 &00 58 06.93&-72 19 17.60&16\,217& PWN & yes & Blue\\
&IKT\,21&1& 57993.47 &01 03 27.99&-72 03 41.30&16\,217& SNR & yes & Blue\\
&1E0102-723&1& 57993.47 &01 03 27.99&-72 03 41.30&16\,217& SNR & yes & Blue\\
&SNR\,C2&1& 57993.66 &00 56 31.02&-72 15 48.40&14\,000& SNR & no & Cyan\\
&B0058-71.8&9& 57993.66 &01 00 19.47&-71 28 15.00&14\,000& SNR & yes & Cyan\\ 
&IKT\,25&10& 57993.66 &01 07 23.94&-72 06 41.50&14\,000& SNR & yes & Cyan\\
2017--08--29&XMMU\,J0056.6-7208&1& 57994.55 &00 56 37.69&-72 09 01.40&16\,223& SNR & no & Magenta\\
&N\,S66D&1& 57994.55 &00 56 37.69&-72 09 01.40&16\,223& SNR & no & Magenta\\
&HFPK\,334&12& 57994.55 &01 04 25.17&-72 45 33.80&16\,223& SNR & yes & Magenta\\
&SNR\,C9&1& 57994.74 &01 12 39.07&-73 28 15.40&15\,487& SNR & no & Yellow\\
2017--09--11$^{e}$&MCSNR~J0127-7332&1& 58007.44 &01 27 37.66&-73 35 20.50&16\,224& SNR & yes & Red\\
&XMMU\,J0049.0-7306&1& 58007.63 &00 49 54.48&-73 04 09.70&12\,319& SNR & no & Green\\
&DEM\,S5&13& 58007.63 &00 41 24.39&-73 39 25.40&12\,319& PWN & yes & Green\\
&B0050-72.8&3& 58007.63 &00 51 35.18&-72 36 02.30&12\,319& SNR & yes & Green\\
2017--09--12&SNR\,C3&1& 58008.54 &01 03 27.99&-72 03 41.30&16\,221& SNR & no & Blue\\
&IKT\,16&7$^{d}$& 58008.54 &00 58 06.93&-72 19 17.60&16\,221& PWN& yes & Blue\\
&IKT\,21&1& 58008.54 &01 03 27.99&-72 03 41.30&16\,221& SNR & yes & Blue\\
&1E0102-723&1& 58008.54 &01 03 27.99&-72 03 41.30&16\,221& SNR & yes & Blue\\
&SNR\,C2$^{f}$&1& 58008.73 &00 56 31.02&-72 15 48.40&5\,436& SNR & no & Cyan\\
&B0058-71.8&9& 58008.73 &01 00 19.47&-71 28 15.00&5\,436& SNR & yes & Cyan\\ 
&IKT\,25&10& 58008.73 &01 07 23.94&-72 06 41.50&5\,436& SNR & yes & Cyan\\
2017--09--28&XMMU\,J0056.6-7208&1& 58024.52 &00 56 37.69&-72 09 01.40&16\,218& SNR & no & Magenta\\
&N\,S66D&1& 58024.52 &00 56 37.69&-72 09 01.40&16\,218& SNR & no & Magenta\\
&HFPK\,334&12& 58024.52 &01 04 25.26&-72 45 33.30&16\,218& SNR & yes & Magenta\\
&SNR\,C9&1& 58024.71 &01 12 39.07&-73 28 15.40&13\,469& SNR & no & Yellow\\
2017--12--13&MCSNR~J0127-7332$^{g}$&1& 58100.47 &01 27 37.66&-73 35 20.50&17\,852& SNR & yes & Red\\

\hline
\end{tabular}
\begin{tablenotes}
\item $^{\textit{a}}$The observation Epoch refers to the start of an observation.
\item $^{\textit{b}}$The PMB coordinates within which the SNR was located.
\item $^{\textit{c}}$$t_{int}$ is the observation integration time.
\item $^{\textit{d}}$IKT\,16:  The PMB was rotated by 60\textsuperscript{o}, resulting in a different beam number corresponding to the same beam coordinates.  Thus on 2017--08--28 IKT\,16 was located in beam 2, but in beam 7 on 2017--09--12.
\item $^{\textit{e}}$During the observations on 2017--09--11 strong RFI was detected.
\item $^{\textit{f}}$SNR\,C2:  The observation was terminated due to strong winds.
\item $^{\textit{g}}$MCSNR~J0127-7332:  Additional Director's Time was allocated due to severe RFI during the 2017--09--11 pointing.
\end{tablenotes}
\end{table*}

Our observations were carried out between 25 August 2017 and 13 December 2017, using a central frequency of 1382\,MHz with a bandwidth of 400\,MHz, split into 1024 channels, and a temporal resolution of 64\,$\mu s$.  Each field was observed twice (except for the \text{MCSNR~J0127-7332} pointing which was observed 3 times), separated on average by 20 days, with an integration time of $\sim$15\,000\,s per pointing.  This would allow us to cross match candidates identified in a pointing, as well as detect binary pulsars at a different orbital phase. 

\subsection{Survey Sensitivity}\label{sensitivity}

To compare our targeted survey with the previous blind surveys \citep{Crawford2001,Manchester2006} we calculate the average, limiting flux density at 1400\,MHz (S$_{1400}$) for each survey with the survey parameters recorded in Table~\ref{tab:sensitivity}.  The average, fundamental limiting flux density for a survey is given by the radiometer equation

\begin{equation}
    S_{lim} \text{ = } \frac{\sigma\beta T_{sys}}{G\sqrt{n_pt_{int}\Delta\nu}}\,,
\label{eq:flux}
\end{equation}

\noindent where the detection threshold, $\sigma$ was chosen to be 8, and $\beta$\,=\,1.5 for BPSR's 2-bit digitisation, which takes into account instrumental imperfections.  For the PMB the mean system temperature is \text{ }$T_{sys}$\,=\,23\,K \citep{McMahon2008, Keith2010}, while the central PMB's gain is \text{ }$G$\,=\,0.735\,K\,Jy\textsuperscript{-1}, with the number of polarisations, \text{ }$n_p$\,=\,2.  The integration time ($t_{int})$) in seconds, and bandwidth ($\Delta\nu$) in MHz for the various surveys are recorded in Table~\ref{tab:sensitivity}.  We used the same parameter values as outlined in \citet{ridley2013} to calculate our sensitivity limits.  The fundamental limiting flux density given by equation~\eqref{eq:flux} for the \citet{Crawford2001} and \citet{Manchester2006} survey was 0.17\,mJy, while our survey was more sensitive with a limiting flux density of 0.11\,mJy.

\begin{table}
\centering
\caption{Observational setup of SMC radio pulsar surveys, as well as the limiting flux densities at 1400\,MHz for $P$\,$\geq$\,50\,ms.  The flux densities were calculated as outlined in Section~\ref{sensitivity}}
\label{tab:sensitivity}
\begin{tabular}{llllll}
\hline
Survey&Pointings&T$_{int}$&t$_{samp}$&$\Delta\nu$&S$_{1400}$\\
&&(s)&($\mu$s)&(MHz)&(mJy)\\
\hline
\hline
\citet{Crawford2001} & NA & 8\,400 & 250 & 288 & 0.066 \\
\citet{Manchester2006} & 73 & 8\,400 & 1\,000 & 288 & 0.067 \\
This work & 12 & 15\,000 & 64 & 400 & 0.039\\
\hline
\end{tabular}
\end{table}

These limits do not take into account the pulse width or broadening effects that must be accounted for, considering that a survey is not equally sensitive to all periods, pulse shapes, and DMs.  We followed the methodology of \citet{Manchester2001} to model the frequency response of a pulsar.  A series of pulses with frequency $f_p$\,=\,$1$/$P$, where $P$ is the pulse period, is portrayed in the Fourier domain by the fundamental and 8 harmonics.  The respective harmonics have an amplitude of $1$/$S_{lim}$, which are multiplied by a sequence of functions representing the transformation the data will be subjected to.   The first function accounts for the intrinsic pulse shape, which is modelled by a Gaussian with a width of $W_{50}$\,=\,$0.05P$:

\begin{equation}
    g_1\text{(}f_p\text{)} \text{ = } exp\left(-\frac{(\pi f_pW_{50})^2}{4ln(2)}\right).
    \label{eq:gauss}
\end{equation}

\noindent The interstellar medium will then disperse the pulses, and cause a dispersion delay across a bandwidth of $\Delta\nu$, centred at a frequency of $\nu$ given by

\begin{equation}
\tau_{DM} \text{ = } 8.3\times10^3DM\Delta\nu\nu^{-3} s.
\end{equation}

\noindent Thus, $g_2$($f_p$) is modelled by a similar Gaussian as in equation~\eqref{eq:gauss}, but with $W_{50}$ replaced with $\tau_{DM}$.  The final transformation has to account for the effect the finite sampling time ($t_{samp}$) has on the pulses, which is characterised by

\begin{equation}
    g_3\text{(}f_p\text{)} \text{ = } \left|\frac{sin\text{(}\pi f_pt_{samp}\text{)}}{\pi f_pt_{samp}}\right|.
\end{equation}

\begin{figure}
  \includegraphics[width=\columnwidth]{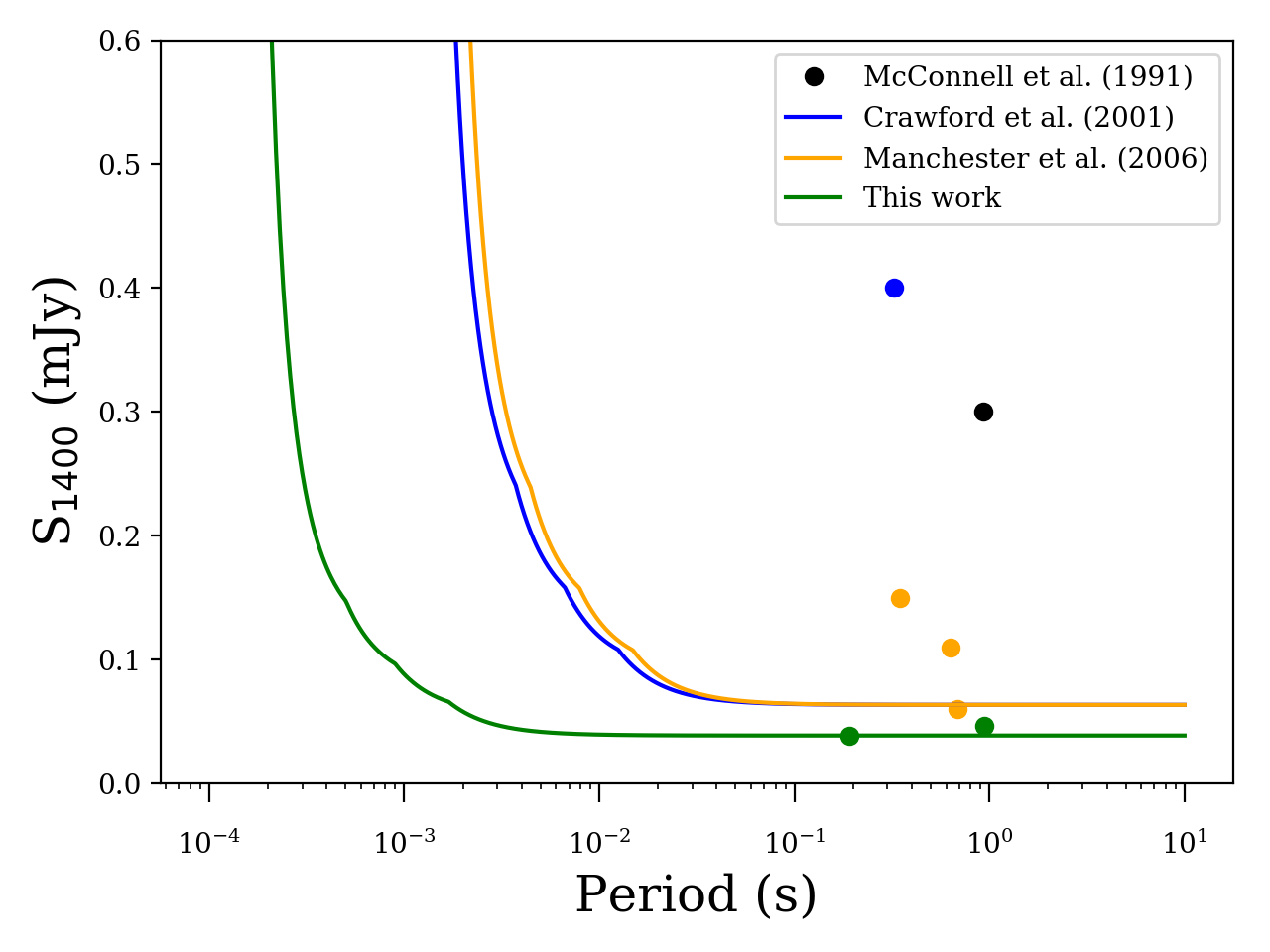}
  \caption{The 1400\,MHz limiting flux densities for all SMC radio pulsar surveys.  The filled circles represent the pulsars detected in the various surveys, in particular the green filled circles indicate the pulsars discovered in this survey (\text{PSR~J0043-73}, \text{PSR~J0052--72}).}
  \label{fig:sensitivity}
\end{figure}

\noindent The final limiting sensitivity is given by summing over $n$ harmonics, where $n$\,$\in$ \,\{1,\,2,\,4,\,8\}.  Figure~\ref{fig:sensitivity} shows the final limiting flux density for the various surveys at 1400\,MHz for the average SMC DM of 116.4\,pc\,cm\textsuperscript{-3}.  It clearly shows our survey is the first to be sensitive to MSPs, and that for $P$\,$\geq$\,50\,ms our survey is twice as sensitive as the previous surveys.  The respective flux densities for $P$\,$\geq$\,50\,ms are recorded in Table~\ref{tab:sensitivity}.

\section{Data Reduction}

\subsection{Young Pulsar Search}\label{sec:yps}

The PMB data were searched for young pulsars (i.e. those with periods of a few tens of milliseconds) using routines from \texttt{PRESTO}\footnote{https://www.cv.nrao.edu/~sransom/presto/} \citep{ransom2002}.  Radio frequency interference (RFI) masks were created for each beam using \texttt{rfifind} with the \texttt{blocks} parameter set to 30.  We searched 3\,672 DMs within a range of 0 $-$ 660 pc\,cm\textsuperscript{-3} (Table~\ref{tab:ddplan}), with no restrictions on the period parameter space.  The dedispersed time series was barycentred to correct for the Earth's motion and then Fourier transformed with \texttt{realfft}.  The resulting power spectra were searched for significant pulses, which were summed up to the 8\textsuperscript{th} harmonic using \texttt{accelsearch} with z\,=\,0, i.e. no acceleration searches were conducted.  Finally, candidates above a S/N threshold of 4.0 were selected and folded with \texttt{prepfold}.  The low threshold was chosen, since we can cross-match candidates between pointings of the same field.  We also observed \text{PSR~J0024-7204C} at the start of every observing day, apart from the 2017--08--29 observations.  The test pulsar was recovered with blind \texttt{PRESTO} searches at an expected S/N of 22, for every observation except on the 2017--09--11, when the observation was subjected to severe RFI.

The search identified $\sim$150\,000 candidates, which were are inspected by eye.  Promising candidates were cross-checked with the alternate pointing of the same source on another day.  If the candidate was not detected in the alternate pointing, the raw data was also folded with the identified period and DM.    

\begin{table}
    \centering
    \caption{Dedispersion parameters for young pulsar search.}
    \begin{tabular}{lllll}
    \hline
  Low DM & High DM & DM step &  Down & Number \\
  (pc\,cm\textsuperscript{-3}) & (pc\,cm\textsuperscript{-3}) & (pc\,cm\textsuperscript{-3}) & Sampling & DMs \\
  \hline
  \hline
    0.00 & 196.80 & 0.10 & 1 & 1968\\
  196.80 & 350.40 & 0.20 & 2 & 768 \\
  350.40 & 580.80 & 0.30 & 4 & 768 \\
  580.80 & 664.80 & 0.50 & 8 & 168 \\
  \hline
    \end{tabular}
    
    \label{tab:ddplan}
\end{table}

\subsection{Longer Period Pulsars}

The Fast Fourier Transform method employed by PRESTO for the young pulsar search is susceptible to rednoise, which reduces the sensitivity for pulsars with periods greater than 200\,ms \citep{VanHeerden2016}.  To combat the effects of rednoise we utilised a fast folding algorithm (FFA, \citealt{Staelin1969}; Morelllo et al. 2019, in prep) to search for pulsars with periods from 200\,ms $-$ 360\,s, and DMs from 0 $-$ 400\,pc\,cm\textsuperscript{-3} with DM steps of 2\,pc\,cm\textsuperscript{-3}.  The lower period limit was selected since the known pulsar distribution peaks at 200\,ms, while the upper limit was the largest period we expect to have sufficient pulses to detect a pulsar.  The data was dedispersed using \texttt{PRESTO}, while applying the same RFI masks created for the young pulsar search.  

The FFA identified $\sim$40\,000 candidates, which were also inspected by eye.  Acceptable candidates were cross matched with the alternate pointing in question, and folded with \texttt{PRESTO} using the period and DM identified by the FFA.

\section{Results}

We discovered two new pulsars, and identified 2 additional low signal to noise (S/N) candidates in our targeted survey of the SMC.  We also detected the known \text{PSR~J0131--7310} \citep{Manchester2006}.  We folded the raw data with \texttt{dspsr}, and then refined the periods, DMs and S/Ns with \texttt{pdmp}.  The quoted values were obtained from \texttt{pdmp}, unless specified otherwise.   Table~\ref{tab:new_pulsar} lists the pulsar properties and the PMB beam coordinates within which the pulsars were detected.

\subsection{PSR~J0131--7310}

\begin{figure}
  \centering
  \includegraphics[width=0.6\columnwidth]{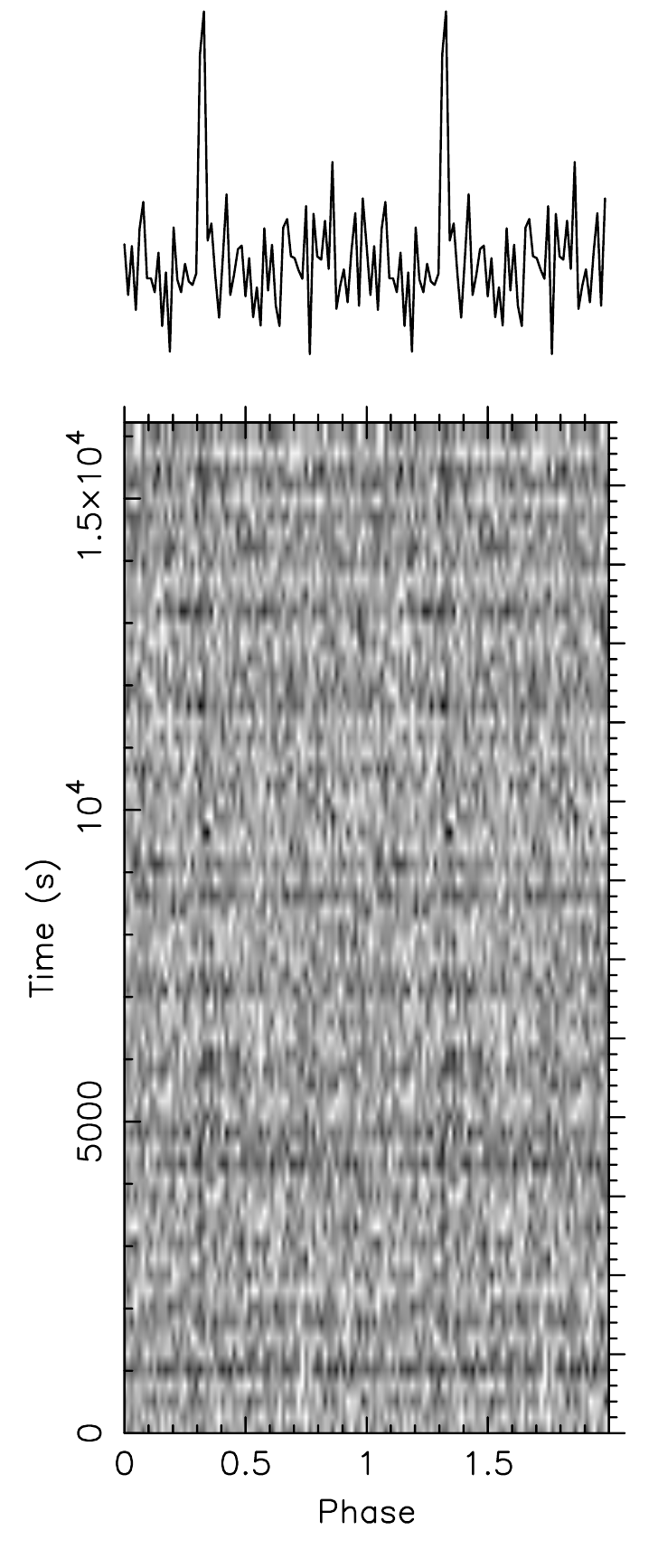}
  \caption{The integrated pulse profile (top panel) of \text{PSR~J0131--7310}, as well as the pulse intensity as a function of pulse phase and integration time (grey scale).  The pulses are plotted over two pulse phases.}
  \label{fig:j0131-7310}
\end{figure}

\text{PSR~J0131--7310} was detected in the 2017--08--25 pointing, focused on \text{MCSNR~J0127-7332}.  The pulsar was on the edge of one of the inner ring beams of the PMB (Figure~\ref{fig:survey}, red square), with beam coordinates of \text{RA\,=\,01:32:46.87} and \text{Dec\,=\,--73:16:21.80} (beam 5).   We detected \text{PSR~J0131--7310} with a barycentric period of 348.12341\,(18)\,ms, and a DM of 206.7\,(1.5)\,pc\,cm\textsuperscript{-3}, matching well the \citet{Manchester2006} values for the period of 348.124045581\,(7)\,ms and a DM of 205.2\,pc\,cm\textsuperscript{-3}.  \text{PSR~J0131--7310} was not detected in the second observation on 2017--09--11, but was detected again in the follow-up observation on 2017--12--13 with the \texttt{FFA}.  Figure~\ref{fig:j0131-7310} shows the pulse profile of \text{PSR~J0131--7310} for the 2017--08--25 pointing.

\subsection{PSR~J0043--73}

\begin{figure*}
  \centering
  \includegraphics[width=\textwidth]{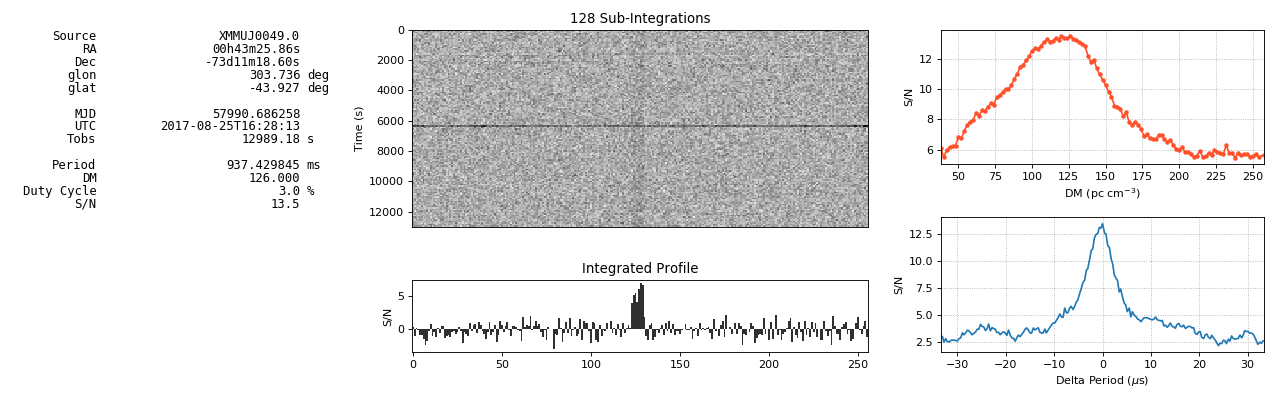}
  \caption{The FFA discovery plot of \text{PSR~J0043--73}, which was located in beam seven of the \text{XMMU\,J0049.0-7306} pointing on 2017--08--25.  The bottom central panel show the integrated pulse profile.  Following the discovery, we folded the raw data with \texttt{pdmp}, constraining the barycentric period to 937.42937\,(26)\,ms and the DM to 115.1\,(3.4)\,pc\,cm\textsuperscript{-3}.  The pulsar was not detected in the initial young pulsar search with \texttt{PRESTO}.}
  \label{fig:new_pulsar}
\end{figure*}

The first new SMC radio pulsar has a barycentric period of 937.42937\,(26)\,ms and DM of 115.1\,(3.4)\,pc\,cm\textsuperscript{-3}, which is consistent with the known SMC pulsar DM range \text{(76\,--\,205\,pc\,cm\textsuperscript{-3})}.  Hereafter the new pulsar will be referred to as \text{PSR~J0043--73} (Figure~\ref{fig:survey}, green square).  The pulsar was found in the 2017--08--25 pointing of which the \text{XMMU\,J0049.0-7306} source was the focus, with beam coordinates of \text{RA\,=\,00:43:25.86} and \text{Dec\,=\,--73:11:18.60} (beam 7), but not in the following observation on 2017-09-11.  The pulsar was discovered using the FFA, but could clearly be seen in the \texttt{PRESTO} \texttt{prepfold} plot when folding the raw data with the identified period and DM.  When we increased the number of summed harmonics to 16, \texttt{PRESTO} also identified \text{PSR~J0043--73} as a candidate pulsar.   Figure~\ref{fig:new_pulsar} shows the FFA discovery plot.  We used the archival Parkes data of the \citet{Manchester2006} survey, and used \texttt{prepfold} to fold the data at \text{PSR~J0043--73}'s period and DM, but did not recover the pulsar.

The known SMC pulsar, \text{PSR~J0045--7319} (Figure~\ref{fig:survey}, black circle on left-hand corner edge of green square) has a similar period, but is located 33.24\,arcmin from the centre of the beam within which \text{PSR~J0043--73} was found.  This is nearly 5 times the HPBW radius, moreover \text{PSR~J0045--7319} has a period of 926.2\,ms and a DM of 105.4\,pc\,cm\textsuperscript{-3} (\citealt{Crawford2001};\citealt{Manchester2005}\footnote{http://www.atnf.csiro.au/research/pulsar/psrcat/}).  Thus, we are confident that \text{PSR~J0043--73} is a new SMC radio pulsar.

\subsection{PSR~J0052--72}

\begin{figure*}
  \centering
  \includegraphics[width=\textwidth]{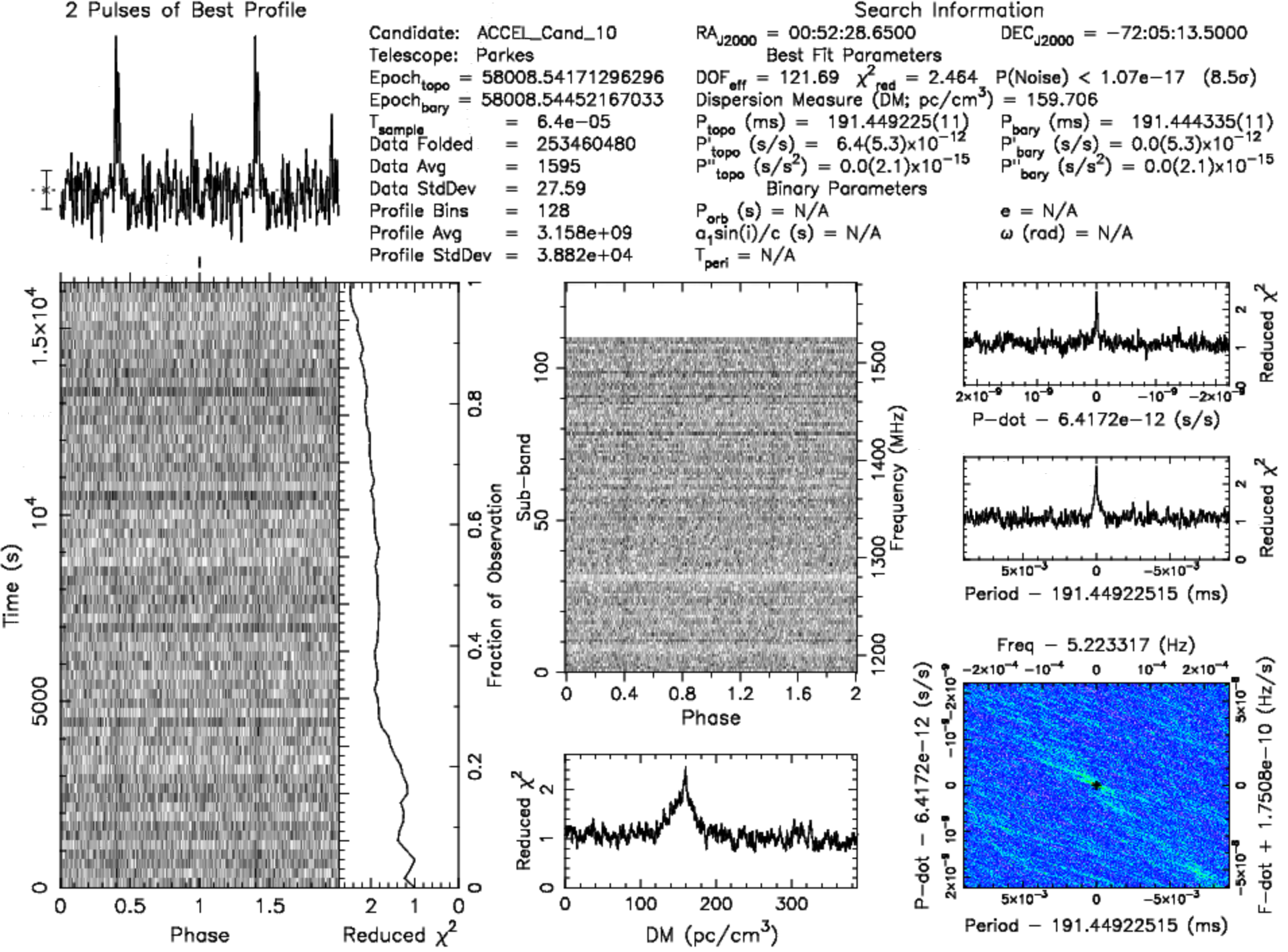}
  \caption{The \texttt{PRESTO} discovery plot of \text{PSR~J0052--72}, which was located in beam eight of the SNR\,C3 pointing on 2017--09--12.  \text{PSR~J0052--72} has a barycentric period to 191.444328\,(46)\,ms and a DM of 158.6\,(1.6)\,pc\,cm\,\textsuperscript{-3}.}
  \label{fig:new_pulsar2}
\end{figure*}

\text{PSR~J0052--72} (Figure~\ref{fig:survey}, blue square) is the second new SMC pulsar (Figure~\ref{fig:new_pulsar2}) with a barycentric period of 191.444328\,(46)\,ms and a DM of 158.6\,(1.6)\,pc\,cm\,\textsuperscript{-3}.  The pulsar was detected in one of the outer ring beams (beam 8) of the 2017--09--12 data set, which was focused on SNR\,C3.  The beam (RA\,=\,00:52:28.65, Dec\,=\,-72:05:13.5) was intersected by two inner ring beams from XMMU\,J0056.6-7208 and SNR\,C2, but \text{PSR~J0052--72} was not detected in either of the beams.  The FFA was adapted to search for the pulsar's period, and subsequently detected the pulsar as well.  \text{PSR~J0052--72} was re-detected in the 2017--08--28 dataset (beam 9) with the FFA.  We also used the archival Parkes data of the \citet{Manchester2006} survey that was coincident with this pointing, and folded the data at the period and DM of \text{PSR~J0052--72}, but the pulsar was not detected in the dataset. 

\subsection{Candidate Pulsars}

We detected two lower S/N candidates, but could not confirm them.  One of the candidates was detected in the 2017--09--12 dataset foucsed on SNR\,C3, and the other in the 2017--09--28 dataset focused on SNR\,C9.  The SNR\,C3 candidate had a period and DM of 21.03561023\,(40)\,ms and 208.42\,pc\,cm\textsuperscript{-3} respectively, while the SNR\,C9 candidate had period of 11.001403386\,(76)\,ms and a DM of 341.58\,pc\,cm\textsuperscript{-3}.  Both candidates were located in beam 5.  We list the candidate pulsars in Table~\ref{tab:new_pulsar}, and show the \texttt{prepfold} plots in Figure~\ref{fig:candidate_snr3} and Figure~\ref{fig:candidate_snr9}.

\section{Discussion}

Two new SMC radio pulsars were discovered in our SMC SNR survey, increasing the SMC pulsar population to seven confirmed pulsars.  Furthermore, two candidates have been identified, but could not be confirmed.  The detected and candidate pulsar properties are shown in Table~\ref{tab:new_pulsar}.

\begin{table*}
    \centering
    \caption{Properties of detected and candidate pulsars discovered in this survey.}
    \begin{tabular}{llllllllll}
    \hline
    \hline
    JName&Date&Beam&Epoch$^{a}$&RA$_{beam}$$^{b}$&Dec$_{beam}$$^{b}$&P$^{c}$&DM&S$_{1400}$&Method\\
    &&&(J2000)&(J2000)&(MJD)&(ms)&(pc\,cm\textsuperscript{-3})&(mJy)&(\texttt{PRESTO}/\texttt{FFA})\\
    \hline
    \text{PSR~J0131--7310}&2017--08--25&5 & 57990.50 & 01 32 46.87 & -73 16 21.80 & 348.12341\,(18)&206.7\,(1.5)&0.157 & FFA\\
    \text{PSR~J0043--73}&2017--08--25 & 7 & 57990.69 & 00 43 25.86&-73 11 18.60 & 937.42937\,(26) & 115.1\,(3.4) & 0.047 & FFA\\
    \text{PSR~J0052--72} & 2017--09--12 & 8 & 58008.54 & 00 52 28.65 & -72 05 13.50 & 191.444328\,(46) & 158.6\,(1.6) & 0.039 & \texttt{PRESTO} \\
    SNR\,C3 candidate & 2017--09--12 & 5 & 58008.54 & 01 09 08.68 & -72 16 34.10 & 21.0356101\,(55) & 208.17\,(66) & 0.038 & \texttt{PRESTO}\\
    SNR\,C9 candidate & 2017--09--28 & 5 & 58024.71 & 01 10 15.74 & -73 55 33.00 & 11.0014027\,(11) & 342.08\,(32) & 0.043 & \texttt{PRESTO}\\
    \hline
    \end{tabular}
    \begin{tablenotes}
    \item $^{\textit{a}}$Observation Epoch is taken to be at the start of a particular observation.
    \item $^{\textit{b}}$The PMB coordinates  within which a pulsar was detected.
    \item $^{\textit{c}}$ This period refers to the barycentric period.
\end{tablenotes}
    \label{tab:new_pulsar}
\end{table*}

\subsection{Detected pulsars}

Our survey schedule enabled us to observe each target twice.  In theory this would allow us to cross-match likely candidates, and so doing confirm candidates as pulsars when they are detected in both observations.  We detected \text{PSR~J0052--72} in both pointings (2017--08--28 and 2017--09--12), however \text{PSR~J0043--73} was only detected in the first pointing on 2017--08--25.  This was a result of more severe RFI, as well as rednoise during the second pointing on 2017--09--11.  In particular, mitigating the rednoise effects for pulsars with periods greater than 200\,ms is not a trivial task when a Fourier based pulsar search is implemented as shown by \citet{VanHeerden2016}.  Thus, it is not suprising that we detected both pulsars, \text{PSR~J0131--7310} and \text{PSR~J0043--73}, with the FFA and not initially in the Fourier based \texttt{PRESTO} search.  Table~\ref{tab:smc_pulsars} lists the properties of the now, seven SMC pulsars.

\begin{table}
    \caption{SMC pulsar parameters.}
    \centering
    \begin{tabular}{lllll}
    \hline
    JName & P & DM & $S_{1400}$ & Discovery\\
    &(ms)&(pc\,cm\textsuperscript{-3})&(mJy)&Ref.\\
    \hline
    \hline
    J0043--73 & 937.42937\,(26) & 115.1\,(3.4) & 0.047 & 1\\
    J0045--7042 & 632.33580002\,(6) & 70\,(3) & 0.11 & 2 \\
    J0045--7319 & 926.27590497\,(3) & 105.4\,(7) & 0.3 & 4 \\
    J0052--72 & 191.444328\,(46) & 158.6\,(1.6) & 0.039 & 1\\
    J0111--7131 & 688.54151164\,(5) & 76\,(3) & 0.06 & 2 \\
    J0113--7220 & 325.88301613\,(1) & 125.49\,(3) & 0.4 & 3 \\
    J0131--7310 & 348.124045581\,(7) & 205.2\,(7) & 0.15 & 2 \\
    \hline
    \end{tabular}
    \begin{tablenotes}
    \item Discovery references:  (1) This work, (2) \citet{Manchester2006}, (3) \citet{Crawford2001}, (4) \citet{mcconnell1991}.
    \end{tablenotes}
    \label{tab:smc_pulsars}
\end{table}

\subsubsection{PSR~J0131--7310}

\text{PSR~J0131--7310} was detected at 20.55\,arcmin from the centre of an inner ring beam, which was coincident with the first side lobe of beam five.  Using the inner ring beam model of \citet{Ravi2016} we scaled the gain appropriately and calculated a flux density of 0.16\,mJy for S/N\,=\,11.90 and a pulse width of 4.20\,ms.  This is comparable to the 0.15\,mJy flux density recorded for \text{PSR~J0131-7310} at 1400\,MHz by \citet{Manchester2006}. 

\subsubsection{PSR~J0043--73}

To calculate the flux density of \text{PSR~J0043--73} we used the radiometer equation with a S/N ratio of 18 and a pulse width of 25\,ms, we determine $S_{1400}$\,=\,0.047\,mJy for an inner ring beam (Figure~\ref{fig:sensitivity}).    We propose it as a new pulsar, in spite of a non-detection in the subsequent observation on 2017--09--11 when the data was compromised with severe RFI. This is also consistent with the non-detection of \text{PSR~J0131-7310} which was also observed on 2017--09--11.  

\subsubsection{PSR~J0052--72}

To calculate the flux density of \text{PSR~J0052--72} we used the radiometer equation with a S/N ratio of 11 and a pulse width of 8\,ms, we determine $S_{1400}$\,=\,0.039\,mJy for an outer ring beam (Figure~\ref{fig:sensitivity}).  This is the fastest spinning radio pulsar discovered in the SMC to date. 

\subsection{Non-detections}

During this survey we observed 15 SNRs (candidates and confirmed), some of which are identified as PWNe.  No pulsars were detected within any of these regions, but we can quantify flux density limits at a 8\,$\sigma$ detection threshold for each of the sources, scaled by their distance from the centre of the beam in question (Table~\ref{tab:snr}).  We used the PMB beam model by \citet{Ravi2016} in combination with equation~\eqref{eq:flux}, and our survey parameters listed in Section~\ref{sensitivity}, which we then multiplied by $\sqrt{\frac{W_{50}}{P-W_{50}}}$, and we set $W_{50}$\,=\,0.05$P$.

\begin{table*}
    \centering
    \caption{Properties of all SNRs/PWNe observed during our survey.  The first six entries was the focus of our survey, which included the known SNR, \text{MCSNR~J0127-7332}, and the five SNR candidates.  The remaining SNRs/PWNe are listed in \citet{Filipovic2008a}.}
    \begin{tabular}{llllllllll}
    \hline
         SNR&Pointing&RA$_{SNR}$$^a$&Dec$_{SNR}$$^a$&RA$_{beam}$$^b$&Dec$_{beam}$$^b$&$D_{extent}$$^c$&$d_{centre}$$^d$&S$_{1400}$$^e$&Type\\
         &&(J2000)&(J2000)&(J2000)&(J2000)&(arcmin)&(arcmin)&(mJy)&(SNR/PWN)\\
         \hline
         \hline
         MCSNR~J0127-7332&MCSNR~J0127-7332&01 27 45.95 &-73 32 56.30&01 27 37.66&-73 35 20.50&-&3.17&0.024& SNR\\
         XMMU\,J0049.0-7306&XMMU\,J0049.0-7306&00 49 46.00&-73 06 17.00&00 49 54.48&-73 04 09.70&0.75&2.21&0.027& SNR\\
         SNR\,C3&SNR\,C3&01 03 35.56&-72 01 35.10&01 03 27.99&-72 03 41.30&1.38&2.18&0.024& SNR\\
         SNR\,C2&SNR\,C2&00 56 25.00&-72 19 05.00&00 56 31.02&-72 15 48.40&3.72&3.31&0.026& SNR\\
         XMMU\,J0056.6-7208&XMMU\,J0056.6-7208&00 56 26.00&-72 09 42.00&00 56 37.69&-72 09 01.40&4.02&1.12&0.024& SNR\\
         SNR\,C9&SNR\,C9&01 12 37.00&-73 26 05.00&01 12 39.07&-73 28 15.40&5.85&2.18&0.025& SNR\\
         \hline
         DEM\,S5&XMMU\,J0049.0-7306&00 41 00.10&-73 36 30.00&00 41 24.24&-73 39 24.80&2.66&3.39&0.035& PWN\\
         B0050-72.8&XMMU\,J0049.0-7306&00 52 36.90&-72 37 18.50&00 51 35.18&-72 36 02.30&2.41&4.79&0.042& SNR\\
         IKT\,16&SNR\,C3&00 58 17.80&-72 18 07.40&00 58 06.93&-72 19 17.60&3.32&1.43&0.036& PWN\\
         IKT\,21&SNR\,C3&01 03 17.00&-72 09 45.00&01 03 27.99&-72 03 41.30&1.20&6.12&0.026& SNR\\
         1E0102-723&SNR\,C3&01 04 01.20&-72 01 52.30&01 03 27.99&-72 03 41.30&0.34&3.14&0.024& SNR\\
         B0058-71.8&SNR\,C2&01 00 23.90&-71 33 41.10&01 00 19.47&-71 28 15.00&3.50&5.45&0.036& SNR\\ 
         IKT\,25&SNR\,C2&01 06 27.50&-72 05 34.50&01 07 23.94&-72 06 41.50&1.13&4.45&0.035& SNR\\
         N\,S66D&XMMU\,J0056.6-7208&00 58 00.00&-72 11 01.40&00 56 37.69&-72 09 01.40&3.34&6.61&0.027& SNR\\
         HFPK\,334&XMMU\,J0056.6-7208&01 03 29.50&-72 47 23.20&01 04 25.17&-72 45 33.80&1.00&4.51&0.032& SNR\\
         \hline
    \end{tabular}
        \begin{tablenotes}
    \item $^{\textit{a}}$The SNR coordinates.
    \item $^{\textit{b}}$The PMB coordinates within which the SNR was located.
    \item $^{\textit{c}}$$D_{extent}$: the angular extent of the SNR.
    \item $^{\textit{d}}$$d_{centre}$: The distance from the centre of the SNR to the centre of the particular PMB beam.
    \item $^{\textit{e}}$$S_{1400}$:  The flux density values are calculated based on $d_{centre}$ for $\sigma$\,=\,8.
\end{tablenotes}
    \label{tab:snr}
\end{table*}

\subsubsection{SNR Candidates}

Our deep, sensitive observations of \text{MCSNR~J0127-7332} and five SMC SNR candidates , identified by the XMM-Newton survey (\citealt{Haberl2012}), and optically selected by MCELS \citep{Smith1999} and SALT Fabry-Perot observations, did not reveal any coincident radio pulsars.  The size of each SNR candidate was less than the HPBW of the central PMB, thus the SNR candidates were each observed in their entirety in a single pointing.  The sensitivity limit for each SNR candidate and \text{MCSNR~J0127-7332} is listed in the first six lines of Table~\ref{tab:snr}.

\subsubsection{\text{MCSNR~J0127-7332} and \text{SNR\,C3} Pointings}

The \text{MCSNR~J0127-7332} and \text{SNR\,C3} pointings were coincident with \text{SXP\,1062} and \text{SXP\,1323} respectively.  Both these X-ray pulsars have extraordinarily long spin periods and are potentially associated with SNRs.  In particular, NSs are typically born with spin periods of tens of milliseconds of (e.g. Crab and Vela pulsars), however \citet{Ikhsanov2012} found that a young NS can possibly be spun down within $10^{4}$\,years by accreting matter from a magnetic wind, if the NS has a magnetic field of $\sim$10$^{13}$\,G.  Alternatively, \citet{Fu2012} as well as \citet{Ho2017} showed that both accreting and isolated NSs with a B\,>\,10\textsuperscript{14}\,G (i.e. a magnetar) can spin down to P\,>\,1000\,s within the lifetime of a SNR.

\vspace{7mm}
\noindent\textit{I) \hspace{3mm} MCSNR~J0127-7332 Pointing}
\vspace{2mm}

\noindent SXP\,1062 is a transient Be\textbackslash X-ray binary (BeXB) with an orbital period of 668 days, a lower limit eccentricity of 0.4, and a X-ray spin period of 1086\,s \citep{Gonzalez-Galan2018}.  SXP\,1062 has been spinning down from a period of 1062\,s in 2010 to 1086\,s in 2014.  This is the second longest spin period known for a BeXB.  If MCSNR~J0127-7332 is the natal SNR of SXP\,1062 it implies a kinematic age of \text{$\sim$2-4$\times$10\textsuperscript{4}\,years} for the NS \citep{Henault-Brunet2012}.  This contradicts standard models, which suggests that NSs at this age are too young to enter an accretion phase, and as a result cannot be X-ray pulsars \citep{Lipunov1992}.  SXP~1062 is in some aspects similar to the highly eccentric gamma-ray binary system \text{PSR~B1259--63}, which has a young radio pulsar of P\,$\sim$\,47\,ms \citep{Johnston1992} with a characteristic age of  \text{3.32$\times$10\textsuperscript{5}\,years} in orbit around a Be star.  The system has a similarly long orbital period of 1236.7\, days, and an eccentricity of 0.87 \citep{Wang2004}.  We searched \text{MCSNR~J0127-7332} for radio pulsations at the X-ray spin period of SXP\,1062, but did not detect any pulsations.  The limiting flux density of this pointing was 0.024\,mJy (Table~\ref{tab:snr}).

\vspace{7mm}
\noindent\textit{II) \hspace{3mm} SNR\,C3 Pointing}
\vspace{2mm}

\noindent SXP\,1323 is another peculiar BeXB with a 26.2 day orbital period, and a X-ray spin period of 1100\,s \citep{Carpano2017}.  SXP\,1323 has the longest spin period known for any accreting X-ray pulsar, but has been rapidly spun up in a period between 2006 and 2016 from 1340\,s to 1100\,s.  Recently, \citet{Gvaramadze2019} found evidence for a putative SNR centred on SXP\,1323.  They identified the SNR with optical studies, however it is the same SNR discovered by XMM-Newton and referred to as \text{SNR\,C3} in this paper.  If the SNR is associated with SXP\,1323 then the NS's age is estimated to be \text{4$\times$10\textsuperscript{4}\,years}, similar to SXP\,1062's age.  This would be the second, long period X-ray pulsar located in its natal SNR, and interestingly both are hosted in the low metallicity environment of the SMC.   However, SXP\,1323 has a much shorter orbital period than SXP\,1062, and subsequently could accrete matter more often.  Accretion typically inhibits radio pulsations, and indeed we did not detect any radio pulsations at the X-ray spin period.

Furthermore, the SNR\,C3 pointing included 2 confirmed SNRs (1E0102-723, IKT\,21) in the central beam, as well as IKT\,16 in the second beam.  1E0102-723 was first detected in the X-rays by \citet{Seward1981}, and then identified by \citet{Dopita1981} as an oxygen rich SNR with narrow band optical imaging.  In a subsequent study, \citet{Vogt2019} identified an isolated neutron star with characteristics consistent with a central compact object (CCO).  From their study the CCO is likely associated with 1E0102-723, thus one would not expect to dectect radio pulsations from 1E0102-723, since CCOs are radio quiet.  Conversely, IKT\,16 has been confirmed as a PWN with a hard X-ray point source \citep{Maitra2015}.  In particular, the X-ray spectral index of 1.1 implies that the emission is dominated by non-thermal emission originating from a pulsar.  The expected spin down power of 10\textsuperscript{37}\,erg\,s\textsuperscript{-1} suggests the existence of a $\sim$100\,ms pulsar.  Down to the 1400\,MHz flux limit of 0.024\,mJy we did not detect any pulsations, thus the pulsar must be either too faint or not beaming towards us.

A pulsar BeXB system, \text{SAX~J0103-722} with a X-ray spin period of 345\,s \citep{Hughes1994,Israel2000,VanderHeyden2004}, is coincident with IKT\,21.  However, it is not clear whether SAX~J0103-722 is associated with IKT\,21 or merely a chance coincidence.  \citet{Hughes1994} argued that a spatial chance coincidence in the SMC is highly unlikely, and would require a transverse spatial velocity exceeding 100\,km\,s\textsuperscript{-1} if the NS of \text{SAX~J0103-722} was born by the same supernova explosion that resulted in the formation of IKT\,21.  Such a large spatial velocity for a BeXB is improbable.  Instead, the study suggests the SN explosion occurred in an OB association which also harbours \text{SAX~J0103-722}.  Nonetheless, we did not detect any radio pulsars down to a flux limit of 0.026\,mJy.

\subsubsection{XMMU\,J0049.0-7306 Pointing}

The \text{XMMU\,J0049.0-7306} pointing included DEM\,S5 and \text{B0050--72.8} (Table~\ref{tab:snr}).  The nature of the compact object originating from the precursor SN explosion is not known for \text{B0050--72.8}, nor has any candidates been identified.  We did not detect any radio pulsations down to a sensitivity of of 0.042\,mJy.  Alasberi et al. 2019 (in prep) has confirmed DEM\,S5 as a PWN, with a morphological structure analogous to \text{PSR~B1951+32} in SNR CTB\,80 \citep{Safi1995} and 'the mouse' \citep{Camilo2002}.  Although this discovery confirms the likely existence of a radio pulsar, we did not detect any radio pulsations down to a flux density of 0.035\,mJy. 

\subsubsection{SNR\,C2 Pointing}

The SNR\,C2 pointing included \text{B0058-71.8} and IKT\,25 in the outer ring beams of the PMB (Table~\ref{tab:snr}).  \text{B0058-71.8} has not been studied in detail, and we did not detect any radio pulsations at a flux limit of 0.036\,mJy.  Conversely, IKT\,25 was likely produced by a thermonuclear (Type Ia) supernova \citep{Roper2015,Lee2011}, thus one would not expect a NS to form, and thus one would not expect a radio pulsar residing in the SNR, and indeed we did not detect radio pulsations down to a sensitivity of 0.035\,mJy.

\subsubsection{XMMU\,J0056.6-7208 Pointing}

N\,S66D was located on the edge of the central beam during the XMMU\,J0056.6-7208 pointing, while HFPK\,334 was positioned in one of the outer ring beams of the PMB.  N\,S66D is poorly studied, consequently no additional information is available in the literature.  On the contrary HFPK\,334 is a young ($\leq$\,1800\,years) \citep{Crawford2014}, intriguing SNR emitting X-ray and radio emission, but not detectable in the optical \citep{Payne2007}.  \citet{Filipovic2008a} noted a central point source within HFPK\,334, and suggested that it may be indicative of a PWN.  In light of their study \citet{Crawford2014} conducted a X-ray study of the central object, but could not identify it as either a CCO or a pulsar, thereupon they proposed that the central point source is a background object.  The lack of any radio pulsations in our study down to a flux limit of 0.032\,mJy, supports their theory that the central source is a background object.  

\section{Conclusions}

We report the first pulsar discoveries in the SMC since the \citet{Manchester2006} survey.  We discovered \text{PSR~J0043--73}, a 937.42937\,ms pulsar with DM\,=\,115.1\,pc\,cm\textsuperscript{-3}, as well as the fastest spinning pulsar to date in the SMC, \text{PSR~J0052--72} with P\,=\,191.444328\,ms and DM\,=\,158.6\,pc\,cm\textsuperscript{-3}.  These discoveries increases the SMC radio pulsar population to seven, corresponding to a 40\% increase in the population.  To date no radio pulsar SNR association has been made in the SMC.

\section*{Acknowledgements}
The Parkes radio telescope is part of the Australia Telescope which is funded by the Commonwealth of Australia for operation as a National Facility managed by CSIRO.  NT, VMc, and DAHB acknowledges support of the National Research Foundation of South Africa (grants 98969, 93405, and 96094).  BWS, VM and MC acknowledge funding from the European Research Council (ERC) under the European Union's Horizon 2020 research and innovation programme (grant agreement No. 694745).

%%%%%%%%%%%%%%%%%%%%%%%%%%%%%%%%%%%%%%%%%%%%%%%%%%

%%%%%%%%%%%%%%%%%%%% REFERENCES %%%%%%%%%%%%%%%%%%

\bibliographystyle{mn2e}
\bibliography{ref_updated.bib}

%%%%%%%%%%%%%%%%%%%%%%%%%%%%%%%%%%%%%%%%%%%%%%%%%%

%%%%%%%%%%%%%%%%% APPENDICES %%%%%%%%%%%%%%%%%%%%%

\appendix

\section{Pulsar candidates}

Here follows the \texttt{prepfold} plots for two pulsar candidates found during the young pulsar search with \texttt{PRESTO.}

\begin{figure*}
  \centering
  \includegraphics[width=.84\textwidth]{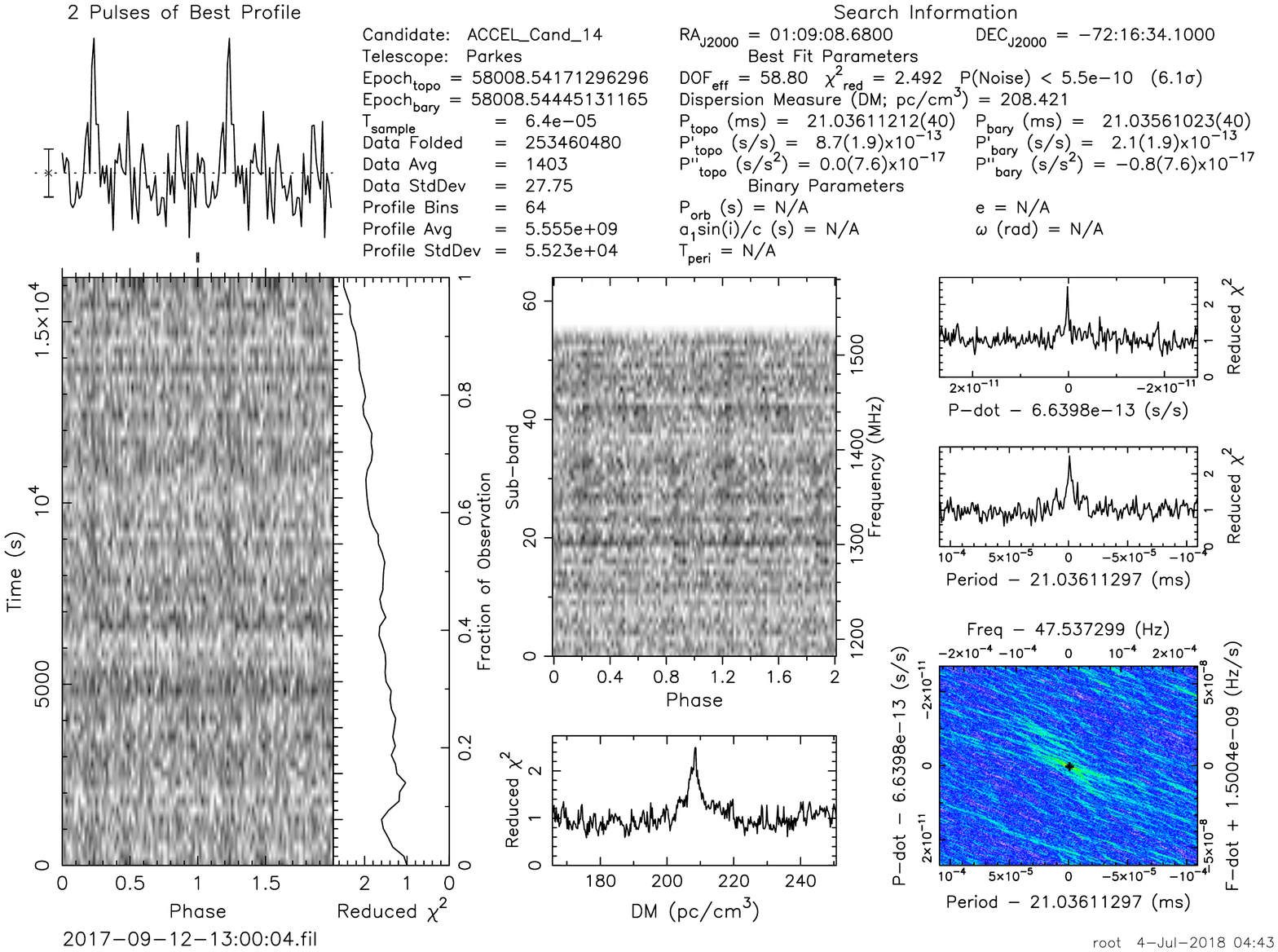}
  \caption{This candidate was detected during a SNR\,C3 pointing on 2017--09--12.  The candidate was located in beam five.}
  \label{fig:candidate_snr3}
\end{figure*}

\begin{figure*}
  \centering
  \includegraphics[width=.84\textwidth]{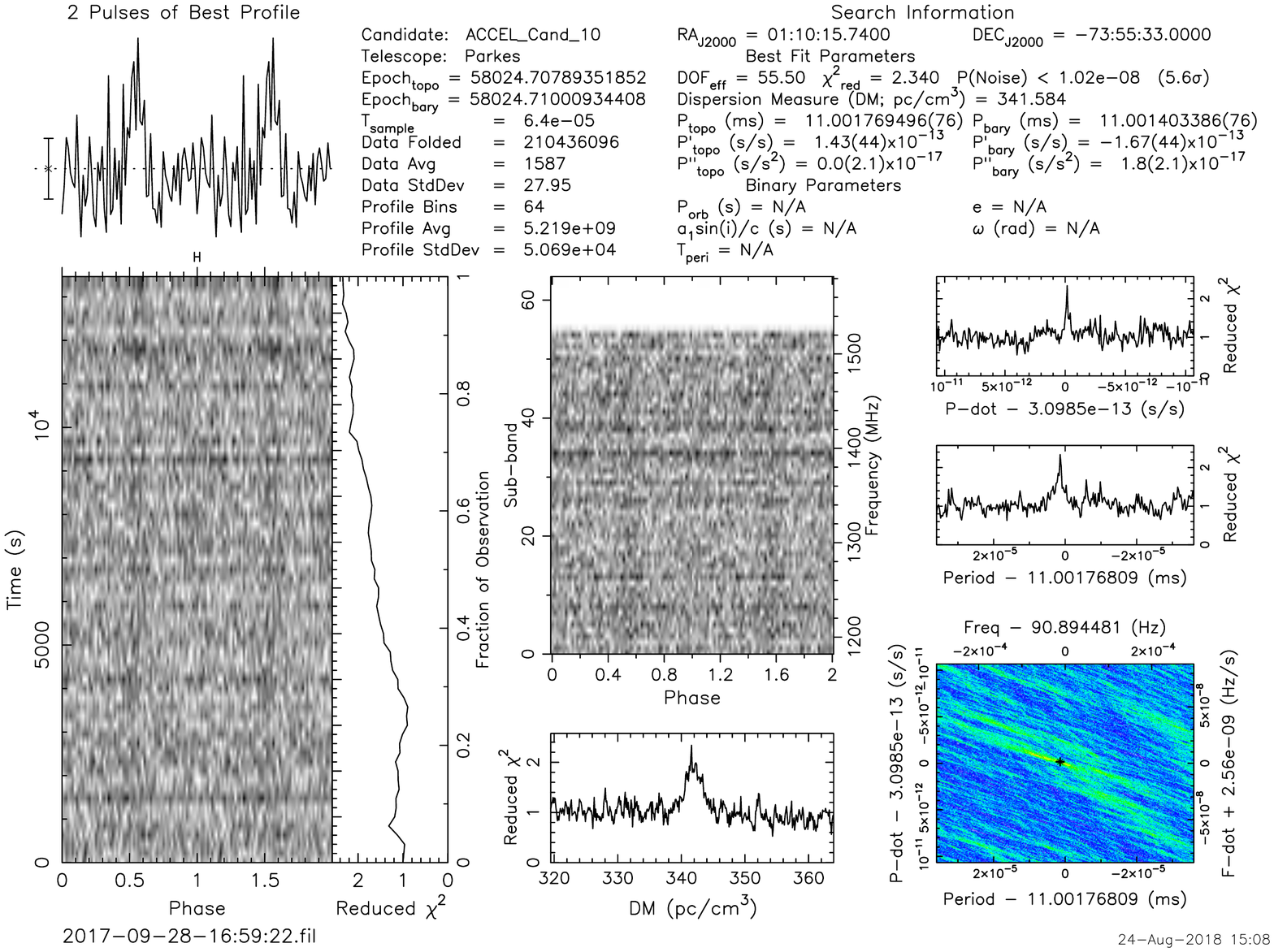}
  \caption{This candidate was detected during a SNR\,C9 pointing on 2017--09--28.  The candidate was found in beam five.}
  \label{fig:candidate_snr9}
\end{figure*}

% Don't change these lines
\bsp	% typesetting comment
\label{lastpage}
\end{document}